\documentstyle[sprocl,epsfig]{article}

\def\Journal#1#2#3#4{{#1} {\bf #2}, #3 (#4)}
\def\Journall#1#2#3#4#5{{#1} #2 {\bf #3}, #4 (#5)}

\def\PRp{\em Phys. Rep.}
\def\NPA{{\em Nucl. Phys.} A}

\def\PLB{{\em Phys. Lett.}  B}
\def\IJMP{{\em Int. J. Mod. Phys.} A}
\def\MPL{{\em Mod. Phys. Lett.} A}

\def\PRD{{\em Phys. Rev.} D}
\def\ZPC{{\em Z. Phys.} C}
\def\PTE{\em Prib. Tekhn. Eksp.}
\def\YaF{\em Phys. At. Nucl.}
\def\ib{\em ibid.}
\def\ea{\em et al.}

\def\ve{\varepsilon}
\def\vp{\varphi_q}

\def\la{\lambda_q}
\def\dq{d_q}
\def\be{\begin{equation}}
\def\ee{\end{equation}}
\def\bea{\begin{eqnarray}}
\def\eea{\end{eqnarray}}

\begin{document}

\begin{flushright}
TAUP 2370-96\\
October 1996
\end{flushright}
\vspace*{2.cm}

\begin{center}
{\large \bf ON DYNAMICS OF FRACTALITY IN CENTRAL C-CU COLLISIONS
AT 4.5$A$ GEV/$c$}
\bigskip
\bigskip

E.K.G. SARKISYAN \\
\smallskip

{\it School of Physics \& Astronomy,
Tel Aviv University,\\ Tel Aviv 69978, Israel}
\medskip
\medskip

L.K. GELOVANI,  G.L. GOGIBERIDZE \\
\smallskip

{\it Joint Institute for Nuclear Research, P.O.B. 79,\\
Moscow 101000, Russia}
\vspace*{1.4cm}

ABSTRACT
\end{center}
\medskip

{\small Fractal structure of charged particle distributions in
pseudorapidity 
and transverse momentum in central C-Cu collisions at 4.5 $A$ 
GeV/$c$ is studied by means of intermittency approach and 
multifractal 
analysis. The modifications to take into account statistical bias
are applied.
Intermittency study in the pseudorapidity phase space indicates 
a 
non-thermal phase transition and different regimes of multiparticle
production. Multifractality is observed within the both methods applied.
The interrelation of the methods      
using   the effective average multiplicity approach 
is studied. 
The findings support 
the idea of statistical significance  of  influence of finite
multiplicities.   
In the transverse momentum spectrum
no dynamical fluctuations are found.} 
\vspace*{1.4cm}

\begin{center}
Talk given at

the 7th International Workshop on Multiparticle Production 

``Correlations and Fluctuations''

Nijmegen, The Netherlands, June 30 -- July 6, 1996
\end{center}

\newpage
\pagestyle{plain}
\setcounter{page}{1}

\title{ON DYNAMICS OF FRACTALITY IN CENTRAL C-CU COLLISIONS
AT 4.5$A$ GEV/$c$}

\author{ E.K.G. SARKISYAN }

\address{School of Physics \& Astronomy,
Tel Aviv University,\\ Tel Aviv 69978, Israel}

\author{ L.K. GELOVANI,$^a$  G.L. GOGIBERIDZE \footnote
{On leave from  Institute of Physics, Tbilisi 380077, Georgia.}}

\address{Joint Institute for Nuclear Research, P.O.B. 79,\\
Moscow 101000, Russia}


\maketitle\abstracts{
Fractal structure of charged particle distributions in  pseudorapidity 
and transverse momentum in central C-Cu collisions at 4.5 $A$ 
GeV/$c$ is studied by means of intermittency approach and 
multifractal 
analysis. The modifications to take into account statistical bias
are applied.
Intermittency study in the pseudorapidity phase space indicates 
a 
non-thermal phase transition and different regimes of multiparticle
production. Multifractality is observed within the both methods applied.
The interrelation of the methods      
using   the effective average multiplicity approach 
is studied. 
The findings support 
the idea of statistical significance  of  influence of finite
multiplicities.   
In the transverse momentum spectrum
no dynamical fluctuations are found. 
}


\section{Introduction}

Last decade investigations of the intermittency effect excited  an enhanced
interest.\cite{rev1}
It is shown \cite{pesch,bnp} that this phenomenon can be a direct
characteristic of a possible phase transition  expected
to be occurred in high energy collisions.

In this talk we present the  results  of  a study of  dynamics of
multiparticle fluctuations (correlations) in relativistic nuclear
collisions 
using the intermittency/fractality approach. Such
investigations have a specific interest due to the expectation of
quark-gluon plasma formation in high energy nuclear collisions and in a
sense of a possible ``soft origin'' of 
intermittency  and weakness of the effect with reaction ``complexity'' and
energy increase.
Note that the study presented continues our preceding  
investigations.\cite{my1,my2,my3}

Dynamical
fluctuations are indicated by a power-law,

\be
{\cal F}_q\, \propto f_q\, M^{\vp},
\qquad 0<\vp\leq q-1
\qquad (q\geq 2),
\label{fi}
\ee
of the $q$th order normalized scaled factorial moments
(NSFM),

\be
{\cal F}_q= M^{q-1}\sum_{m=1}^{M}
\frac{\langle n_m^{[q]}\rangle}
{\langle n_m\rangle ^q}\, ,
\label{fm}
\ee
over $M$ number of the bins into which the
phase space of produced particles is divided.
Here  $n_m$  is  the number of particles in the $m$-th bin  for  an
each event,  $\langle \cdots \rangle $  denote averaging  over the
events, and $n^{[q]}=n(n-1)\cdots (n-q+1)$.

The intermittency indices $\vp$
characterize (multi)fractal structure of the distribution.
Fractality is pointed out \cite{pesch,bnp} to be a reflection of the  
phase
transition via the codimensions defined as

\be
d_q=\vp/(q-1).
\label{dq}
\ee
Monofractal patterns [$d_q ={\rm const.}(q)$] are associated with
second-order phase transition (e.g. from  quark-gluon
plasma),
while self-similar cascading is characterized by multifractals
[$d_q>d_p$ at $q>p$] with a possible ``non-thermal'' phase transition.

As a signal of the transition an existence of a minimum of the function
\be
\la=(\vp+1)/q
\label{lam}
\ee
at a certain ``critical'' value of $q=q_c$ is argued. The minimum of
Eq. \ref{lam} may be also a manifistation of coexistence of
(many small) liquid-type fluctuations and  dust-type (high
density) ones.

To continue  the study to noninteger \footnote{Recently 
the procedure of  continuing
the NSFM to noninteger $q$'s has been suggested.\cite{hwa:ni}}
$q$'s and
to make a direct    search for  multifractality,
the
method of   frequency moments   was proposed.
\cite{hwa:fr} The moments are defined as  
\be
{\cal G}_q= \sum_{m=1}^{M}\left( \frac{n_m}{n}\right) ^q\theta(n_m-q)\, . 
\label{gq}
\ee
Here $n$ is the number of particles per event and $\theta(x)$ is a 
step function which is  1 if $x\geq 0$ and 0 otherwise.

The dynamics of the fluctuations observed is of  self-similar nature, 
i.e. 
\be
\langle {\cal G}_q\rangle \, \propto \, g_q\, M^{-\tau_q}\, ,
\label{tau}
\ee
and in analogous to Eq. \ref{fi} 
 indicates dynamical fluctuations:
\be
\tau_q^{dyn}\approx  q-1- \vp\, .
\label{taufi}
\ee

Let us note that in our previous study \cite{my1} of the ${\cal
G}$-moments 
multifractality of the $\eta$-spectra has been  found 
with indication  of the  dynamical fluctuations. 

\section{Data Sample and Analysis}

The  data   analyzed  here   come   from   interactions  of  the JINR
Synchrophasotron (Dubna)  4.5~$A$ GeV$/c$  $^{12}$C beam  with a  copper
target  inside  the 2m  Streamer  Chamber  SKM-200.\cite{skm}
The   central
collision trigger was used:   absence of charged particles with  momenta
$p>3$ GeV$/c$ in a forward cone of 2.4$^{\circ }$ was  required.

The scanning and handling of the  film data were carried out on  special
scanning tables  of the  Lebedev Physical  Institute (Moscow).\cite{obr}
The
average       measurement       error       in       the        momentum
$\langle\ve_p/p\rangle$ was about 12$\%$, and that in the  polar
angle  measurements  was   $\langle\ve_{\vartheta}\rangle\simeq
2^{\circ}$.   A total  of 305  events with  charged  particles in  the
pseudorapidity window $\Delta \eta=0.2-3.0 $ are considered  ($\eta=-\ln
\tan  (\vartheta/2)$).  The  accuracy $\langle\ve_{\eta}\rangle$
does  not  exceed  0.1.  In
addition   particles  with  $p_t>1$  GeV/$c$  are  excluded from the
investigation to  eliminate the  contribution of  protons as  far as  no
negative  charged  particles  were  observed  with  such
transverse momentum in an
assumption of equal numbers of positive and negative pions.  The  average
multiplicity is of $21.2\pm 0.6$.

The  fluctuations  are  considered  in  the $\eta$ and 
$p_t$ phase spaces.  To avoid the problem  of
the non-flat shape  of the distributions  $\rho (x), x=\eta\, {\rm or}
  \ln p_t$
 we use a ``cumulative'' variable,\cite{fl1,fl2}

\begin{equation}
\stackrel{\sim }{x}\; =
\int_{x_{min}}^{x}
\rho(x')dx' \,
/
\int_{x_{min}}^{x_{max}} \rho(x')dx'\; ,
\label{nv}
\end{equation}
with  the 
uniform spectrum $\rho(\stackrel{\sim }{x})$  within the [0,1] interval.

Further, we apply  the modified SFM  (MSFM)
proposed \cite{mm}  to  remove the biased
estimator  of  the  normalization of the NSFM (\ref{fm})
(especially in small  bins):  
\begin{equation}
F_q =
\frac{{\cal N}^q}{M}
\sum_{m=1}^{M}
\frac{\langle n_m^{[q]}\rangle }
{N_m^{[q]}
}
\; .
\label{fb}
\end{equation}
Here $N_m$ is the number of particles in the $m$th bin in all $\cal  N$
events.

The modified scaled frequency moments (MSFrM) were  suggested
\cite{blaz:pc,blaz:mm} to be defined as
\be
G_q=
\frac{{\cal N}}{M}
\sum_{m=1}^{M}
\frac{\langle n_m^q \, \theta(n_m-q)
\rangle }
{N_m^q}\, .
\label{mgq}
\ee

Later on 
the modified moments (\ref{fb}) and (\ref{mgq}) instead of the biased
ones (\ref{fm}) and (\ref{gq}) are studied. 
The scaling laws (\ref{fi}) and (\ref{tau}) are considered with
respect to these unbiased quantities. 
Note that use of the ``transformed'' variable (\ref{nv}) along with the
modifications   allow to study higher-order moments.

\section{Results}
\subsection{Intermittency in the Pseudorapidity Phase Space}
\label{fmpseudo}

Fig.  1   illustrates  dependence  of the $F_q$ 
(\ref{fb}) on $M$, depicted for $q=3$, $4$ for
pseudorapidity fluctuations. The
different increase of the MSFM with $M$ for the different
$M$-regions
manifesting on the plots 
continues upto $q=8$ (not shown) confirming our earlier
results.\cite{my1}
Such a behavior of the moments
lends support to the existence of distinguished regimes
of particle creation at various bin averaging scales.  
\vspace*{.3cm}
\begin{center}
\begin{tabular}{lr}
\epsfig{file=f3ea.epsi,height=2.2in}
&
\epsfig{file=f4ea.epsi,height=2.2in}
\end{tabular}
\end{center}
\vspace*{.1cm}
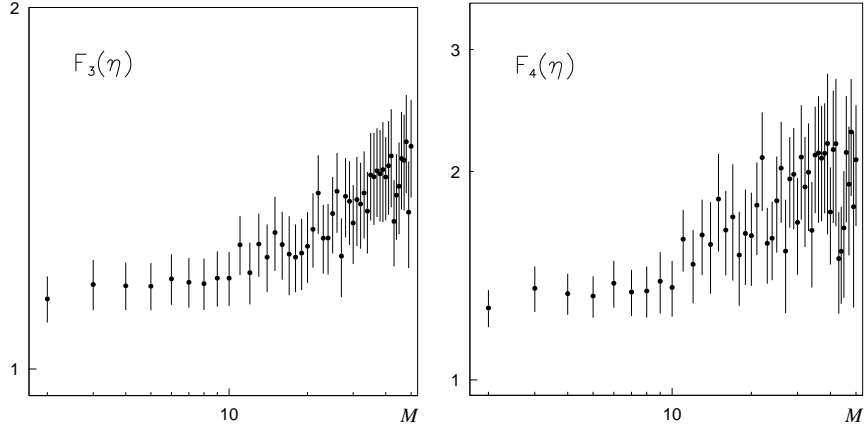
\centerline{\footnotesize{Figure 1: Log-log plots of the MSFM (\ref{fb}) vs. $M$ in
the pseudorapidity phase space.}} 
\begin{center}
\begin{tabular}{lr}
\epsfig{file=dfl.epsi,height=2.2in}
&
\epsfig{file=lamb.epsi,height=2.2in}
\end{tabular}
\end{center}
\vspace{.1cm}
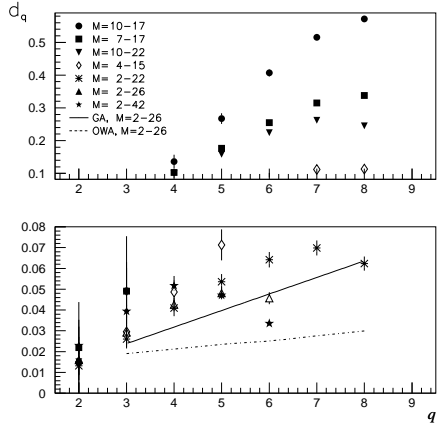
\centerline{\footnotesize{Figure 2: $d_q$ (\ref{dq}) vs.
$q$.~~~~~~~~~~~~~~~~~~~~~~~~~~Figure 3: $\la$ (\ref{lam}) vs. $q$.}}

Fig. 2  presents the function $\dq$ (\ref{dq}) for different
$M$-intervals.  A few intervals with sensitively
distinguished increase of the $\dq$ are seen:
$2\leq M \leq 22$, $4\leq M
\leq 15$, $7\leq M \leq 17$, and $10\leq M \leq 17$.
Multifractality observed, as mentioned above, points out a cascading
scanario of particle production.

From Fig. 3 of displaying of $\la$ function (\ref{lam}) one  can
conclude that at least two regimes of particle  production  exist: 
one with the phase transition at  $q_c =4$, and another one
for which no critical behavior is  reached.  Taking into  account
multifractality,
 critical $q_c$ indicates \cite{pesch}  ``non-thermal''
phase transition.  Although the interpretation may be a matter
of debate the minimum was found earlier in hadronic interactions
\cite{rev1} at small $p_t$ and very recently by means of the $M$-intervals
study in ultra-high heavy ion collisions.\cite{uhc}

In Fig. 2 and 3  we show also the $\la$ predicted by the
Gaussian approximation (GA) \cite{pesch} and Ochs-Wosiek approach (OWA)
\cite{fl2}.
  As seen  the approximations meet difficulties,
especially when $q_c$ exists; the same is found for the negative
binomial distribution input \cite{my3}. Since all these random cascading
models are
based on the second order MSFM, the difference indicates
multiparticle character of a possible phase transition.
This observation confirms our earlier results \cite{my2,my3} despite the
restriction of 
transverse momenta.

\subsection{Intermittency in the Transverse Momentum  Phase Space}
\label{fmpt}

Fig. 4 shows the MSFM behavior with $M$ in the phase space of $p_t$. No
intermittency effect is observed in this projection. The fluctuations in
the transverse
momentum distributions seem to be  of statistical origin.  
\vspace*{0.4cm}
\begin{center}
\begin{tabular}{lr}
\epsfig{file=f2p.epsi,width=5.5cm,height=1.4in}
&
\epsfig{file=f3p.epsi,width=5.5cm,height=1.4in}
\end{tabular}
\end{center}
\vspace*{.1cm}
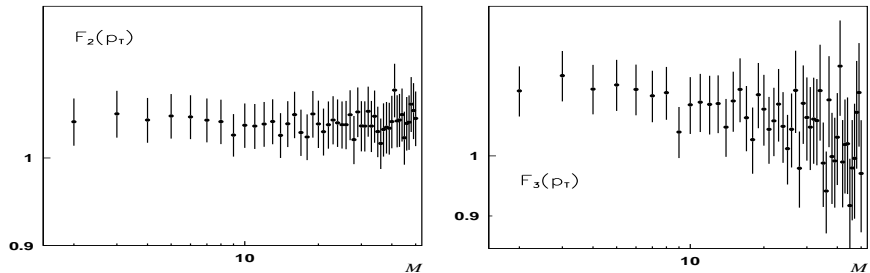
\centerline{\footnotesize{Figure 4: Log-log plots of the MSFM (\ref{fb}) vs. $M$ in
the $p_t$ phase space.}} 
\vspace*{.15cm}

\subsection{Multifractality  Analysis}

The MSFrM (\ref{mgq}) at $q=2, 3$ calculated in the
pseudorapidity phase space as
well as in the
$p_t$-projection are shown in Fig. 5 
in comparison with  the statistical $G_q^{\rm stat}$. A clear
difference of
the pseudorapidity fluctuations from the 
\begin{center}
\begin{tabular}{lr}
\epsfig{file=g2epa.epsi,height=2.3in}
&
\epsfig{file=g4epa.epsi,height=2.3in}
\end{tabular}
\end{center}
\vspace*{.1cm}
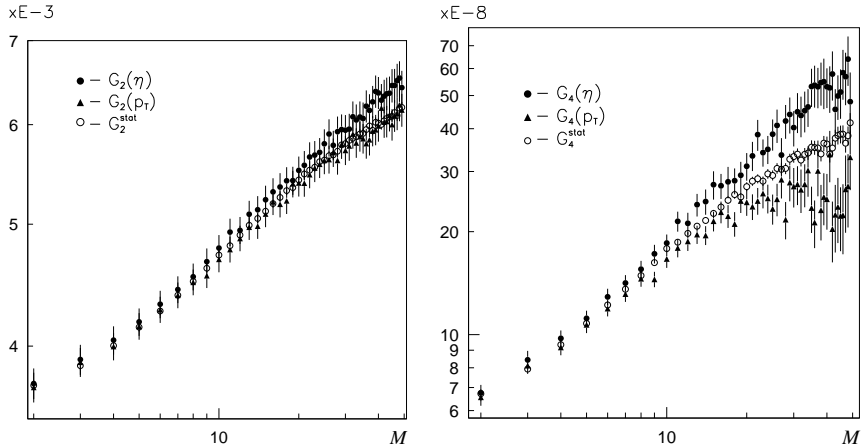
\centerline{\footnotesize{Figure 5: Log-log plots of the MSFrM (\ref{mgq}) vs. $M$.}}
\vspace*{.3cm}
statistical ones is seen, while    
the fluctuations coming from the distributions in the transverse momentum
confirm their statistical nature.

The effect is much  stronger for very small bins (large $M$'s). Taking
into account that  scaling laws (\ref{fi}) and (\ref{tau}) are
satisfied (strictly) for large $M$ (mathematically $M \to \infty$) this
finding seems to be very important and shows real dynamical origin of the
fluctuations in $\eta$-projection. Two different behaviors of
the MSFrM -- weak
rising
at $M\leq 10$ and its enhancement for $M>10$ -- indicate two possible
regimes of
particle production mentioned  in Sec.
\ref{fmpseudo}.

Contrary  to the observed \cite{rev1} decrease and fast
saturation of the
$\cal G$-moments for $q>0$ (as it was also shown
\cite{my1} by
us) the
MSFrM are increasing functions of $M$ upto very small bins. 
Despite this, connected with the modified form (\ref{mgq}), the
strong 
increase indicates very multifractal structure of the $\eta$-spectrum. 

Note that the method of  frequency moments  allows to build
fractal spectral function  which in a case  of the phase transition
should have zeros at $q_c$.\cite{pesch,blaz:pc,blaz:phtr} Let us mention
that    
in our study \cite{my1}  of C-Ne collisions such a zero was observed.

The scaling indexes $\vp$ and $\tau_q$  are  connected
via Eq. \ref{taufi}, 
approximate character of which  
is shown \cite{blaz:eam} to be caused  by influence  of finite
multiplicities. Very recently 
to  solve  the problem 
the
procedure based on the method \cite{blaz:eam}   
of the effective average multiplicity 
has been  developed.\cite{blaz:mm} 
It is proposed to use the generalized moments defined  by
multiplying (in our case) by the $N_m^{-L}, L=0,1,2,\dots $ under the
sum  in the definitions
(\ref{fb}) and (\ref{mgq}), so that for $L=0$ the standard forms are
restored.  

The generalized scaling laws,
$$F_q\, \propto  \left( {\cal N}/M \right) ^{q-1}\,
[N_0^{(F)}]^{-L}f_q\, M^{\vp}\,
,$$


$$G_q\, \propto  [N_0^{(G)}]^{-L}f_q\, M^{\vp}\, ,$$
are pointed out \cite{blaz:pc,blaz:mm} to  characterize dynamical
fluctuations. 
Here $N_0^{(F),(G)}$ is the effective average multiplicity predicted
\cite{blaz:mm} to be independent of the type of the moments 
($N_0^{(F)} = N_0^{(G)}$)  as
well as of the parameter $L$.

\begin{center}
\begin{tabular}{lr}
\epsfig{file=fn0e.epsi,height=2.2in}
&
\epsfig{file=gn0e.epsi,height=2.2in}
\end{tabular}
\end{center}
\vspace*{.1cm}
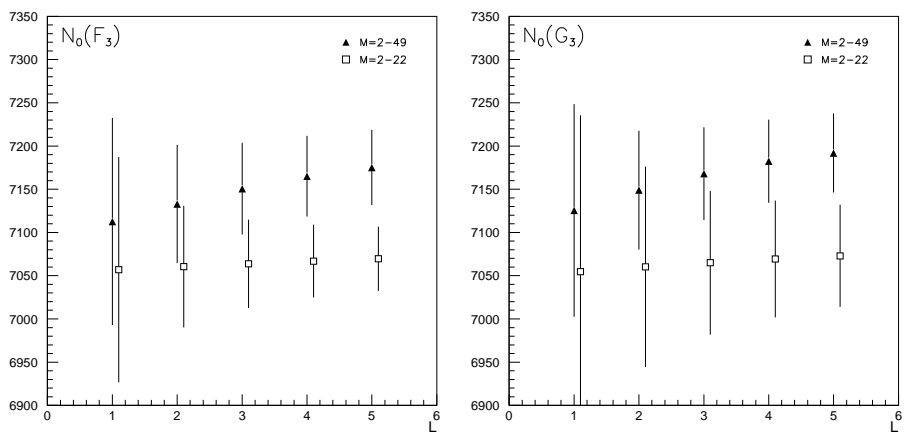
\centerline{\footnotesize{Figure 6: Effective average multiplicity
as a function of $L$.}}
\vspace*{.3cm}
Fig. 6 confirms  the mentioned independences and shows a weak
dependence of $N_0^{(F),(G)}$ on the $M$-interval in both analysis
perfomed.  
These findings gives evidence for real accounting of statistical
contribution (shown \cite{my1,my2} to be sensitive) to find an exact
relation between the slopes. Note in this sense the method \cite{hwa:ni} 
of noninteger order factorial moments where  multifractal analysis is
continued to the intermittency approach.

\section{Conclusions}

Study of dynamics of fluctuations and fractality in the pseudorapidity and
transverse momentum spectra of charged particles produced in central C-Cu
collisions at 4.5 Gev/$c$ is performed. The  scaled factorilal
and frequency moments are calculated in the transformed variables and 
corrected to take into
account the bias of infinite
statistics. The fluctuations in the pseudorapidity phase space are found
to have very multifractal structure and to be of dynamical nature,
indicating possible non-thermal phase
transition. Two different regimes of particle production during the
hadronization cascade are indicated. The generalized scaling laws of the
moments  analyzed 
are studied 
via  the method of the effective average multiplicity. 
Correct relation between the scaling exponents are shown to be
strongly influenced
by  finite multiplicities.     
The fluctuations in the transverse momentum projection 
are found to be 
statistical ones.

\section*{Acknowledgments}

E.S. is grateful to the Organizing Committee and especially to Prof.
W.Kittel for inviting him to participate in this very fruitful  Workshop.
We would like to thank  M.Bla\v{z}ek for his useful cooperation and
interesting suggestions and all participants for their interest to our
work.

\end{document}